\input{aipcheck}

\documentclass[,final]{aipproc}

\layoutstyle{8x11double}

\begin{document}

\title{Ultra High Energy Cosmic Rays Propagation and Spectrum}

\classification{96.50sb 96.50.sd 95.30.Cq 96.50.sh}
\keywords      {Particle Astrophysics, Astrophysical Backgrounds, Cosmic Rays Observations}

\author{Roberto Aloisio}{
  address={INFN - Laboratori Nazionali del Gran Sasso, ss 17bis Km 18+910, Assergi (AQ), Italy}
}

\begin{abstract}
The status of the observations of Ultra High Energy Cosmic Rays will be reviewed, focusing on the  
the latest results of HiRes and Auger observatories. A comprehensive analytical computation
scheme to compute the expected UHECR spectrum on earth will be presented, applying such scheme to 
interpret the experimental results. The phenomenological scenarios favored by HiRes and Auger in terms of 
chemical composition and spectrum will be also presented.
\end{abstract}

\maketitle

\section{Introduction}
Ultra High Energy Cosmic Rays (UHECR) are the most energetic particles observed in nature with 
energies up to $3\times 10^{20}$ eV. The detection of these particles, started already 
in the 50s with the pioneering experiments of Volcano Ranch in the USA and Moscow University array
in the USSR, poses many interesting questions mainly on the origin and chemical composition of such 
fascinating particles. In the recent years a new step forward was done with the measurements performed 
by HiRes and AGASA first, and nowadays with the first five years results of the Pierre Auger Observatory 
in Argentina. 

The propagation of UHECR from the sources to the observer is mainly conditioned by the interaction of 
these extremely energetic particles with the intervening astrophysical backgrounds such as the Cosmic
Microwave Background (CMB) and the Extraglactic Background Light (EBL). The study of the propagation 
of UHE particles through these backgrounds is of paramount importance to interpret the observations and 
to discover the astrophysical origin of UHECR. 
Several propagation dependent features in the spectrum can be directly linked to the chemical composition 
of UHECR and/or to the distribution of their sources \cite{GZK,dip,AloBon}. Among such features particularly 
important is the Greisin, Zatsepin and Kuzmin (GZK) suppression of the flux \cite{GZK}, an abrupt depletion 
of the observed proton spectrum due to the interaction of the UHE protons with the CMB radiation field. 
The GZK suppression, as follows from the original papers \cite{GZK}, is referred to protons and it is due to 
the photo-pion production process on the CMB radiation field ($p+\gamma_{CMB} \to \pi + p$). In the case of 
nuclei the expected flux also shows a suppression at the highest energies that, depending on the nuclei 
specie, is due to the photo-disintegration process on the CMB and EBL radiation fields 
($A+\gamma_{CMB,EBL} \to (A-nN) + nN$). In any case, the interaction processes between UHE particles 
and astrophysical backgrounds will condition the end of the CR spectrum at the highest energies and
the high energy behavior of the flux can be used as a diagnostic tool for the chemical composition of the 
observed particles. 
Another important feature in the spectrum that can be directly linked with the nature of the primary particles and on 
their origin (galactic/extra-galactic) is the pair-production dip \cite{dip}. This features is present only in the 
spectrum of UHE extragalactic protons and, as the GZK, is a direct consequence of the proton interaction 
with the CMB radiation field, in particular the dip brings a direct imprint of the pair production process 
$p+\gamma_{CMB} \to p + e^{+} + e^{-}$ suffered by protons in their interaction with CMB radiation. 

From the experimental point of view the situation is far from being clear with different experiments claiming
contradictory results. The HiRes experiment shows a proton dominated spectrum till the highest energies 
with a clear observation of the protons GZK suppression and the pair-production dip \cite{HiRes}. The chemical
composition observed by HiRes is coherent with this picture showing a proton dominated spectrum at all 
energies $E>10^{18}$ eV \cite{HiRes}. The situation changes if the Auger results are taken into account.
The latest release of the observed Auger spectrum \cite{Auger} shows a not clear confirmation of the 
pair production dip and of the protons GZK suppression. Signaling a possible deviation from a
proton dominated spectrum to an heavier composition at the highest energies. This picture is confirmed by the 
chemical composition observed by Auger, that shows an heavy mass composition at energies 
$E>4\times 10^{18}$ eV \cite{Auger}. 

This puzzling situation, with different experiments favoring different scenarios, shows the importance 
of a systematic study of UHECR propagation in astrophysical backgrounds. In the present paper we 
will review the main points of the kinetic theory of UHECR propagation, comparing theoretical results with 
the latest experimental data of Auger ad HiRes. The paper is organized as follows: in section 2 we review the 
kinetic theory of propagation for protons and nuclei, in section 3 we will discuss different theoretical 
scenarios in comparison with experimental data, conclusions will take place in section 4. 

\section{Protons and Nuclei Propagation}

The propagation of charged particles (protons or nuclei) in astrophysical backgrounds can be suitably studied 
in the kinetic approach, that enables the computation of the expected fluxes on earth given any hypothesis on the 
astrophysical sources of UHECR. The main assumption under which the kinetic theory is build is the 
Continuum Energy Losses (CEL) approximation \cite{BereBook}, through which particle interactions are 
treated as a continuum process that continuously depletes the particles energy. 

In the propagation through astrophysical backgrounds the interactions of particles are naturally affected 
by fluctuations, with a non-zero probability for a particle to travel without loosing energy. In our computation
scheme such fluctuations are neglected; as was shown in \cite{dip,BereKach} this approach has 
a limited effect on the flux computation. Only at the highest energies ($E>100$ EeV) fluctuations produce 
a deviation of the order of $10\%$ respect to the flux computed with a standard Monte Carlo
simulation \cite{BereKach}.

As already anticipated in the introduction UHECR propagating through astrophysical backgrounds suffer 
different interaction processes loosing energy and, in the case of nuclei, being even destroyed.

\begin{itemize}
\item{{\it protons}} - UHE protons interact only with the CMB radiation field giving rise to the two processes of pair
production and photo-pion production. Both of these reactions can be treated in the CEL hypothesis. 

\item{{\it nuclei}} - UHE nuclei interact with the CMB and EBL radiation fields, suffering the process of pair
production, for which only CMB is relevant, and photo-disintegration, that involves both backgrounds CMB 
and EBL. While the first process can be treated in the CEL hypothesis, being the 
nuclei specie conserved, the second produces a change in the nucleus specie. Following \cite{AloBere10} we 
will treat the photo-disintegration process as a "decaying" process that simply depletes the flux of the propagating
nucleus. 
\end{itemize}
 
The propagation of UHE particles in the energy range $E\simeq 10^{18} \div 10^{19}$ eV is extended over
cosmological distances with a typical path length of the order of Gpc. Therefore we should also take into account
the adiabatic energy losses suffered by particles because of the cosmological expansion of the Universe. 

The kinetic equation that describes the propagation of protons and nuclei are respectively: 
\begin{equation}
\frac{\partial n_p(\Gamma,t)}{\partial t} - \frac{\partial}{\partial 
\Gamma} \left [ b_p(\Gamma,t)n_p(\Gamma,t) \right ] = Q_p(\Gamma,t)
\label{eq:kin_p}
\end{equation}   
\begin{equation}
\frac{\partial n_{A}(\Gamma,t)}{\partial t} - \frac{\partial}{\partial \Gamma}
\left [ n_{A}(\Gamma,t) b_{A}(\Gamma,t) \right ] + \frac{n_{A}(\Gamma,t)}
{\tau_{A}(\Gamma,t)}  = Q_{A}(\Gamma,t)
\label{eq:kin_A}
\end{equation}
where $n$ is the equilibrium distribution of particles, $b$ are the energy losses (adiabatic expansion of the Universe
and pair/photo-pion production for protons or only pair-production for nuclei) $Q$ is the injection of freshly
accelerated particles and, in the case of nuclei, also the injection of secondary particles produced by 
photo-disintegration (see below). 

Concerning sources, in this paper, we will consider the simplest hypothesis of a uniform distribution of sources that
inject UHECR of different species with a power law injection of the type: 
\begin{equation}
Q_{inj}(\Gamma, z) = Q_0(z) \Gamma^{-\gamma_g} e^{-\Gamma/\Gamma_{max}}
\label{eq:injection}
\end{equation}
where $\Gamma$ is the Lorentz factor of the particle, $\Gamma_{max}$ is the maximum Lorentz factor that the
sources can provide and $\gamma_g$ is the power law index of the distribution of the accelerated particles at the
sources. In our discussion, depending on the particle specie, we will consider different injection parameters $Q_0$
and $\Gamma_{max}$, keeping the power law index $\gamma_g$ fixed for all species.

The energy losses $b$ for protons or nuclei depend only on the CMB field and in the CEL hypothesis can be
computed analytically through \cite{dip}:
\begin{equation}
b (\Gamma) = \frac{c}{2\Gamma} \int_{\epsilon_0} d\epsilon_r \sigma(\epsilon_r)f(\epsilon_r)\epsilon_r
\int_{\epsilon_r/2\Gamma} d\epsilon \frac{n_{CMB}(\epsilon)}{\epsilon^2}
\label{eq:loss1}
\end{equation}
where $\Gamma$ is the Lorentz factor of the UHE particle, $\epsilon_r$ is the energy of the background photon in the 
rest frame of the UHE particle, $\epsilon_0$ is the threshold of the considered reaction in the rest system of the UHE
particle, $\sigma(\epsilon_r)$ is the cross-section of the process, $f(\epsilon_r)$ is the mean fraction of energy lost by
UHE particle in a single collision in the laboratory system, and $n_{CMB}(\epsilon)$ is the density of the CMB
background photons in the Laboratory reference frame. 

The energy losses (\ref{eq:loss1}) do not take into account the expansion of the Universe and are intended at the
present epoch $(z=0)$, in order to generalize energy losses at any red-shift $z$, because in the computation of $b$ 
only CMB plays a role, one can simply use the following recipe $b(\Gamma,z)=(1+z)^2 b((1+z) \Gamma,z=0)$.
The expansion of the universe is taken into account by adding $b(\Gamma,z)+\Gamma (1+z) H(z)$, with 
$H(z)=H_0\sqrt{(1+z)^3 \Omega_m + \Omega_{\Lambda}}$. In the present paper we will always assume standard
cosmology with $H_0=72$ Km/(s Mpc), $\Omega_m=0.27$ and $\Omega_{\Lambda}=0.73$.

\begin{figure}
  \includegraphics[height=.2\textheight]{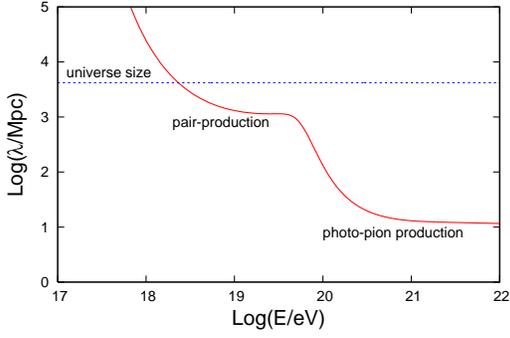}
  \caption{protons path length, with labels of the relevant process that fix the behavior in 
  different energy regimes. The sharp suppression of the path length responsible of the expected GZK 
  cut-off is clearly seen as a consequence of the photo-pion production process. Only the CMB radiation
  field is relevant for proton propagation (see the text).}
  \label{fig1}
  \end{figure}

In figure \ref{fig1} we show the behavior of the path-length $\lambda=c/b_p$ (at $z=0$) for protons, from the 
figure it is evident the effect of the photo-pion production process with a sharp decrease of the contributing 
Universe that is at the basis of the expected GZK cut-off in the proton spectrum.

\begin{figure}
  \includegraphics[height=.21\textheight]{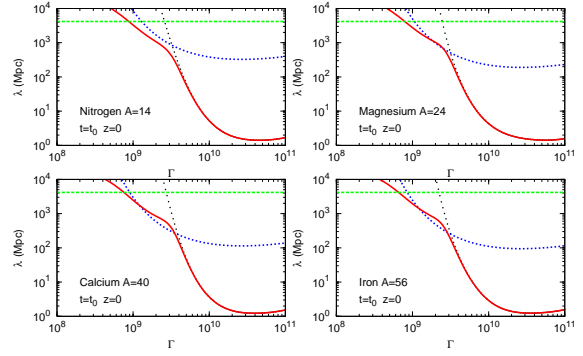}
  \caption{Nuclei path length under the processes of pair production $\lambda=c/b_A$ (blue dotted) and 
  photo-disintegration $\lambda=c\tau_A$. In the case of photo-disintegration we have considered 
  the path-length computed taking into account all relevant backgrounds CMB and EBL (red continuos line)
  and only CMB (black dotted line). The size of the Universe is also shown as greed dashed line.}
  \label{fig2}
  \end{figure}

In the case of nuclei the energy losses term $b_A$ in equation (\ref{eq:kin_A}) depends only on the pair-production 
process on CMB and on the universe expansion. The rate $b_A$ can be computed as in equation (\ref{eq:loss1})
and it is easy to show that it is related to the protons pair-production loss rate by the relation:
\begin{equation}
b_A(\Gamma)= \frac{Z^2}{A} b_b(\Gamma)
\label{eq:pair_A}
\end{equation}
where $Z,A$ are the charge (atomic number) and atomic mass number of the nucleus. The second process 
that affects nuclei propagation is photo-disintegration over CMB and EBL backgrounds. This process
is treated as a decaying process that depletes the flux of nuclei, it enters in the kinetic equation (see equation 
(\ref{eq:kin_A})) through a sort of "life-time" of the nucleus under the photo-disintegration process. This 
"life-time" corresponds to the mean time needed to a nucleus of Lorentz factor $\Gamma$ and atomic number 
$A$ to lose, at least, one of its nucleons: 
\begin{equation}
\frac{1}{\tau_A}=\frac{c}{2\Gamma^2}
\int_{\epsilon_0(A)}^{\infty} d\epsilon_r \sigma(\epsilon_r,A)\nu(\epsilon_r)\epsilon_r
\int_{\epsilon_r/(2\Gamma)}^{\infty} d\epsilon \frac{n_{bkg}(\epsilon)}{\epsilon^2}
\label{eq:loss2}
\end{equation}
where $\sigma(\epsilon_r,A)$ is the photo-disintegration cross-section and $\nu(\epsilon_r)$ is 
the molteplicity associated to this process, namely the average number of nucleons extracted 
from the nucleus by a single interaction and $n_{bkg}=n_{CMB}+n_{EBL}$. 
The dependence on red-shift of $\tau_A$ directly follows from the evolution with red-shift of the background 
photon densities $n_{CMB}$ and $n_{EBL}$. In the case of CMB this dependence is known analytically 
while for the EBL we will refer to the models presented by Stecker \cite{Stecker_EBL}. In figure \ref{fig2} we 
show the behavior of the path length $\lambda$ for different nuclei species, being $\lambda=c/b_A$, for 
pair-production, or $\lambda=c\tau_A$, for photo-disintegration. In figure \ref{fig2} we show also the effect of the 
EBL by plotting $\lambda$ computed taking into account both backgrounds (continuos line) and only CMB (black 
dotted line). It is important to note that the photo-disintegration process starts to contribute to the propagation of
nuclei at a Lorentz factor that is almost independent of the nuclei specie $\Gamma_{cr}\simeq 2\times 10^{9}$
\cite{AloBere10}. This is an important general characteristic of nuclei photo-disintegration process from which 
we can immediately deduce the dependence on the nucleus specie of the energy corresponding to the 
photo-disintegration suppression of the nuclei flux: 
\begin{equation}
E^A_{cut} = A m_N \Gamma_{cr}
\label{eq:GZK_A}
\end{equation}
being $A$ the atomic number of the nucleus and $m_N$ the proton mass. From equation (\ref{eq:GZK_A}) 
it is evident how the flux behavior could bring informations on the chemical composition of the UHECR,
in the case of Helium ($A=4$) the suppression is expected around energies $E \simeq 10^{19}$ eV 
while in the case of Iron ($A=56$) the suppression is expected at higher energies $E\simeq 10^{20}$ eV. 

Let us discuss now the generation function $Q_A(\Gamma,t)$ in the rhs 
of Eq.~(\ref{eq:kin_A}). One should distinguish among primary nuclei, i.e. nuclei 
accelerated at the source and injected in the intergalactic space, and secondary 
nuclei and protons, i.e. particles produced as secondaries in the photo-disintegration
chain.  
In the case of primaries the injection function has the form 
(\ref{eq:injection}), while the injection of secondaries should be modeled 
taking into account the characteristics of the photo-disintegration process. 
The dominant process of photo-disintegration is the one nucleon ($N$)
emission, namely the process $(A+1) +\gamma_{bkg} \to A+N$, this follows directly 
from the behavior of the photo-disintegration cross-section (see \cite{AloBere10} and 
references therein) that shows the giant dipole resonance corresponding 
to one nucleon emission. Moreover, at the typical energies of UHECR ($E>10^{17}$ eV) 
one can safely neglect the nucleus recoil so that photo-disintegration will conserve 
the Lorentz factor of the particles. Therefore the production rate of secondary 
$A-$nucleus and $A-$associated nucleons will be given by
\begin{equation}
Q_A(\Gamma,z)= Q_p^A(\Gamma,z)=
\frac{n_{A+1}(\Gamma,z)}{\tau_{A+1}(\Gamma,z)}
\label{eq:injA}
\end{equation}
where $\tau_{A+1}$ is the photo-disintegration life-time of the nucleus father $(A+1)$ 
and $n_{A+1}$ is its equilibrium distribution, solution of the kinetic equation
(\ref{eq:kin_A}). 

Using equation (\ref{eq:injA}) we can build a system of coupled differential equations that
starting from primary injected nuclei $(A_0)$, with an injection given by (\ref{eq:injection}), follows 
the complete photo-disintegration chain for all secondary nuclei $(A<A_0)$ and nucleons. Clearly 
secondary protons\footnote{Neutrons decay very fast into protons, so we will always refer to 
secondary protons.} propagation will be described by the proper kinetic equation (\ref{eq:kin_p}) 
with an injection term given by (\ref{eq:injA}). 
The solution of the kinetic equation for protons and nuclei can be worked out analytically. In the case
of protons:
\begin{equation}
n_p(\Gamma,z)=\int_{z}^{z_{max}} \frac{dz'}{(1+z')H(z')} 
Q_p(\Gamma',z) \frac{d\Gamma'}{d\Gamma}
\label{eq:np-solut}
\end{equation}
being $Q_p$ the injection of primary protons (equation (\ref{eq:injection})) or secondary
protons (equation (\ref{eq:injA})) and $\Gamma'=\Gamma'(\Gamma,z)$ is the characteristic function of 
the kinetic equation \cite{AloBere10}. In the case of nuclei:
\begin{equation}
n_A(\Gamma,z)=\int_{z}^{z_{max}} \frac{dz'}{(1+z')H(z')} 
Q_A(\Gamma',z)\frac{d\Gamma'}{d\Gamma} e^{-\eta_{A}(\Gamma',z')}.
\label{eq:nA-solut}
\end{equation}
being, again, $Q_A$ the injection of primary nuclei (\ref{eq:injection}) or secondary
(\ref{eq:injA}). The exponential term in Eq.~(\ref{eq:nA-solut}) is given by 
\begin{equation}
e^{-\eta_{A}(\Gamma ',z ')} = \exp\left [-\int_{z}^{z'} \frac{dz''}{(1+z'')H(z'')}
\frac{1}{\tau_{A}(\Gamma '', z'')}\right ],
\label{eq:etaA}
\end{equation}
in which one easily recognizes the survival probability 
during the propagation time $t'-t$ for a nucleus with fixed $A$.
Therefore, $\exp(-\eta)$ provides the suppression of large $z'$ in 
the integral in Eq.~(\ref{eq:nA-solut}). 
The derivative term $d\Gamma'/d\Gamma$ present in both solutions 
(\ref{eq:np-solut}) and (\ref{eq:nA-solut}) can be computed according to 
\cite{AloBere10}. Finally, the maximum red-shift of integration $z_{max}$ 
can be directly computed solving the equation $\Gamma'(\Gamma,z)=E_{max}$, 
assuming a maximum acceleration energy $E_{max}=A m_N \Gamma_{max}$ 
($A=1$ for protons) in the injection of freshly accelerated particles. 

\begin{figure}
  \includegraphics[height=.3\textheight]{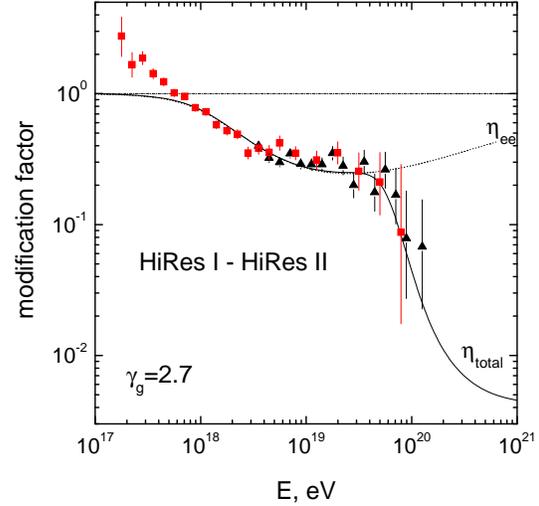}
  \caption{Pure proton spectrum in comparison with HiRes data \cite{HiRes}, data are from 
  HiRes II monocular (boxes) and HiRes I monocular (triangles). The two curves $\eta_{tot}$ and $\eta_{ee}$
  show the total spectrum and the spectrum calculated including only adiabatic and pair-production energy
  losses.}
  \label{fig3}
  \end{figure}

\section{Comparison with Observations}

Using the solutions of the kinetic equation (\ref{eq:np-solut}) and (\ref{eq:nA-solut}) we can compute the 
expected flux on earth (at $z=0$). In this section we will discuss the two alternative scenarios that follows 
from the observations of Auger and HiRes. 
Let us first consider the case of HiRes data. In figure \ref{fig3} we show the comparison of 
these data with a pure proton spectrum, the comparison is performed by means of the modification factor
$\eta(E)$ \cite{dip}. This quantity is given by the ratio of the energy spectrum $n_p(E)$ calculated from equation 
(\ref{eq:np-solut}) (with all energy losses taken into account), and the unmodified spectrum $n_{unm}(E)$, where 
only adiabatic energy loss (Universe expansion) is included: $\eta(E) = n_p(E)/n_{unm}(E)$ \cite{dip}.
The modification factor is a convenient quantity to describe, in a model independent way \cite{dip}, any 
feature in the spectrum related to energy losses. From figure \ref{fig3} it is evident the agreement of the 
HiRes observed spectrum with the expected pair production dip, typical of a pure proton spectrum 
with a best fit injection power law $\gamma_g=2.7$ \cite{dip}.

The nature of the spectrum suppression seen in figure \ref{fig3} can be 
further characterized taking the integral spectrum as observed by HiRes. 
The GZK suppression for protons in the integral spectrum can be tagged by the energy scale $E_{1/2}$, that
corresponds to the energy at which the integral spectrum $J(>E)$ becomes half of the simple power-law
extrapolation spectrum $KE^{-\gamma}$. Theoretical calculations for protons predict a value 
$E_{1/2}=10^{19.72}$ eV for a wide range of injection power laws $\gamma_g=2.1\div 2.8$ \cite{BereE1/2}. 
The measured value of $E_{1/2}$ as claimed by the HiRes collaboration is in fairly good agreement with the 
predicted value, being $E_{1/2}^{HiRes} = 10^{19.73\pm 0.07}$. 
From these observations of HiRes, with some caution, one may conclude that HiRes has detected the two 
main signatures of a pure proton composition: the pair production dip and the GZK suppression.
This important conclusion of the HiRes observations is confirmed from the direct measurement of the UHECR
chemical composition performed by this experiment. In particular, the HiRes measurement of the elongation rate
$X_{max}$ and its root-mean-square (RMS) confirm a strong dominance of protons in the UHECR composition
at all energies $E>10^{18}$ eV \cite{HiRes}. 

The picture emerging from HiRes observations is not confirmed by Auger. The Auger data on mass composition 
and spectra are quite different from those of HiRes. In particular, the Auger observations
show a nuclei dominated spectrum already at energies around $4\times 10^{18}$ eV. This conclusion is quite robust
in the Auger data and it follows from both observations on elongation rate as well as on its RMS. The change 
with energy of mass composition in the Auger data is quite smooth with a progressive increase of the heavy nuclei
content starting already at $2\times 10^{18}$ eV. 

\begin{figure}
  \includegraphics[height=.2\textheight]{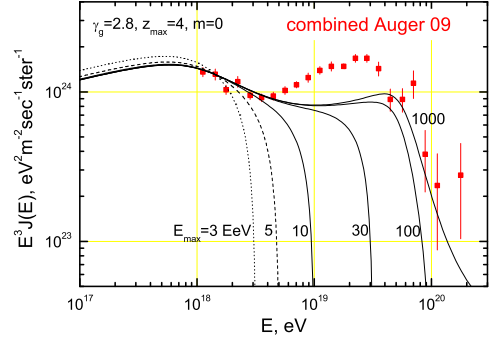}
  \caption{Pure proton spectrum with an injection power law $\gamma_g=2.8$ and for different choices
   of the maximum energy $E_{max}$ (as labelled) in comparison with the observed Auger flux.}
  \label{fig4}
  \end{figure}

Coherently with the elongation rate observations the UHECR flux observed by Auger is difficult to reconcile with a
pure proton spectrum. In figure \ref{fig4} we show the comparison of the Auger spectrum with a pure proton 
spectrum obtained fixing the injection power law $\gamma_g=2.8$ and with different choices of the maximum
attainable energy at the source (as labeled). It is clear that the observed spectrum is well described by a pure 
proton flux only at energies below $(3\div 5) \times 10^{18}$ eV. At larger energies the theoretical proton spectrum 
is not compatible with the Auger observations \cite{disapp}, that don't show the typical dip behavior as discussed 
in the case of HiRes data. The same conclusion holds changing the injection power law, taking a very flat 
injection with $\gamma_g=2.0$ the situation doesn't change with the theoretical proton spectrum compatible 
with Auger observations only at low energy, $E<(3\div 5) \times 10^{18}$ \cite{disapp}.

As discussed in the case of HiRes observations, another important signature of the proton content at the highest
energies is related to the shape of the flux suppression. Also under this respect the flux observed by Auger seems 
not compatible with a pure proton composition. The value of $E_{1/2}$ and the behavior of the suppression are 
better related to an heavy nuclei composition than a pure proton.

\begin{figure}
  \includegraphics[height=.2\textheight]{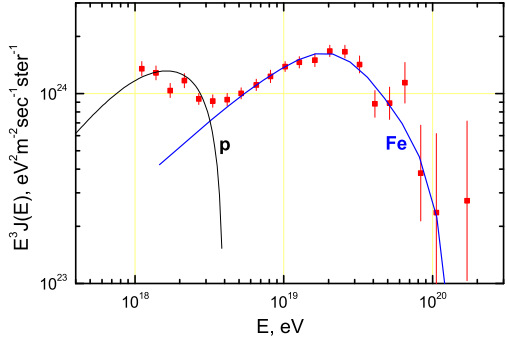}
  \caption{UHECR spectrum in a two component model with protons and Iron nuclei, the injection 
  power law index is $\gamma_g=2.0$ and the maximum energy is $E_{max}=4 Z \times 10^{18}$ eV.}
  \label{fig5}
  \end{figure}

As recently pointed out in \cite{disapp}, the experimental evidence emerging from Auger data can be understood
in terms of different primary species and their maximum energy at the source. A large set of astrophysical
acceleration mechanisms, in particular those based on diffusive shocks, show a maximum energy at the source 
which is rigidity dependent, being related to the electric charge of the accelerated particle:  
$\Gamma^{A}_{max}\propto Z \Gamma^{p}_{max}$ with $\Gamma^{A}_{max}$ maximum Lorentz factor of a
nucleus with atomic number $Z$ and $\Gamma^{p}_{max}$ maximum Lorentz factor for protons. 
Using this characteristic of the maximum energy it is possible to explain the progressively
heavier composition with increasing energy observed by Auger. This goal is reached by the so-called 
"disappointing model" \cite{disapp}, that describes the Auger observations by means of a multi-component injection 
of UHECR at the sources. At energy higher than $Z E_{max}^p$ nuclei with charge $Z' < Z$ disappear, while heavier
nuclei with larger Z survive. Starting from $E^p_{max} \simeq (4 \div 10) \times 10^{18}$ eV, the higher energies are
accessible only for nuclei with progressively larger values of $Z$. In particular, the maximum observed energy must
correspond to Iron nuclei, which can reach $E_{max}^{Fe} \simeq (1\div 3) \times 10^{20}$ eV.
In figure \ref{fig5} we plot the UHECR spectrum in a two component model: protons and Iron, in comparison
with the Auger observed flux, for an injection power law index $\gamma_g=2.0$ and a value of the maximum energy
for protons of $E_{max}^p=4\times 10^{18}$ eV. From this figure it is evident the good agreement of this simple
(two-components) model with the Auger observations. The predictions of this model are very disappointing for future
detectors. In fact, the maximum energy for Iron $E_{max}^{Fe} = (1\div 4)\times 10^{20}$ eV implies an energy 
per nucleon below the threshold $E_N<E_{max}^{Fe}/A\simeq (2\div 5)\times 10^{18}$ eV, i.e. well below the GZK
energy. Therefore, the production of secondary detectable particles such as neutrinos or gamma-rays will be
subdominant respect to the photo-disintegration process of nuclei and no alternative detection of the UHECR 
propagation will be possible \cite{disapp}. Moreover, it will be impossible any correlation study of the UHECR 
events with possible astrophysical sources because the effect of galactic magnetic field will substantially deviate 
the nuclei trajectories at the highest energies \cite{disapp}. For these reasons the model at hand, while gives 
a good description of the Auger observations, was named "disappointing" by the authors of \cite{disapp}. 

\section{Conclusions}

In conclusion, we face at present the most serious disagreement in the observational data of the two
biggest experiments in UHECR. HiRes observes signatures of proton propagation through CMB in the form of 
the pair-production dip and GZK cutoff. Moreover, these observations are well confirmed by the HiRes direct 
measurement of a proton dominated mass composition. The study of UHECR propagation enables us to 
accommodate HiRes observations in a pure proton model, with sources characterized by a steep injection spectrum
$\gamma_g=2.5\div 2.8$ and an high maximum energy $E_{max}^p= (1\div 3) \times 10^{20}$ eV.

On the other hand, Auger clearly observes an high-energy
steepening of the spectrum, but its position and shape are rather different from the prediction of the GZK cutoff.
Moreover, the behavior of the spectrum observed by Auger in the energy range 
$1\times 10^{18} \div 4\times 10^{19}$ eV doesn't confirm the pair-production dip typical of protons, signaling a
substantial nuclei contamination in the flux observations. Coherently with this, the mass composition
directly observed by Auger at $E \ge 4 \times 10^{18}$ eV shows a dominance of
nuclei that becomes progressively heavier increasing the energy and reaches a pure Iron composition at 
$E \simeq 10^{19}$ eV. 

From our analysis follows that the Auger data favor a multi-component
spectrum at the sources with a conservative explanation in terms of flat injection $\gamma_g=2.0\div 2.3$
and a relatively low maximum energy for protons 
$E_{max}^{p}\simeq (3\div 5)\times 10^{18}$ eV, that corresponds to a maximum energy for Iron nuclei 
of the order of $E_{max}^{Fe}\simeq (1\div 2) \times 10^{20}$ eV. This scenario emerging from Auger observations, 
if confirmed, will be quite disappointing because the dominance of nuclei at the highest energies will seriously 
harm any experimental study of correlation with sources as well as any detection of UHE neutrinos or gamma-rays
produced by UHECR propagation. 

Let us conclude by stating that the experimental observation of the UHECR chemical composition at the highest
energies has a paramount importance in choosing among the two alternative scenarios depicted in this paper and 
establishing the future directions of this field of research. Unfortunately the available observations at the highest
energies are still affected by poor statistics and renewed experimental efforts are needed in order to unveil the 
nature of UHECR.

\begin{theacknowledgments}

I'm grateful to V. Berezinsky, A. Gazizov and S. Grigorieva with whom the present
work was done. This paper was partially supported by the Gran Sasso Center for Astroparticle Physics
(CFA) funded by the European Union and Regione Abruzzo under the contract P.O. FSE
Abruzzo 2007-2013, Ob. CRO.

\end{theacknowledgments}

\bibliographystyle{aipproc}

\begin{thebibliography}{9}

\bibitem{GZK} 
K. Greisen, Phys. Rev. Lett. {\bf 16}, 748 (1966);
G.T. Zatsepin and V.A. Kuzmin, Pisma Zh. Experim. Theor. Phys. {\bf 4}, 114 (1966). 

\bibitem{dip}
V. Berezinsky, A. Gazizov and S. Grigorieva, Phys. Rev. D {\bf 74}, 043005 (2006);
R. Aloisio, V. Berezinsky, P. Blasi, A. Gazizov, S. Grigorieva and B. Hnatyk, Astrop. Phys. {\bf 27}, 
76 (2007).

\bibitem{AloBon}
R. Aloisio and D. Boncioli, arXiv:1002.4134 [astro-ph.HE].

\bibitem{HiRes}
HiRes collaboration, Phys. Rev. Lett. {\bf 100} (2008) 101101;
P. Sokolsky,  arXiv:1010.2690 [astro-ph.HE];
HiRes collaboration, Phys. Rev. Lett. {\bf 104} (2010) 161101. 

\bibitem{Auger}
Auger collaboration, Phys. Letters {\bf B 685} (2010) 239;
Auger collaboration, Phys. Rev. Lett. {\bf 104} (2010) 091101.

\bibitem{BereBook}
V. Berezinskii, S. Bulanov, V. Dogiel, V. Ginzburg and V. Ptuskin, 
Astrophysics of Cosmic Rays, North-Holland, 1990. 

\bibitem{BereKach}
V. Berezinsky, A. Gazizov and M. Kachelriess, Phys. Rev. Lett. {\bf 97} (2006) 23110.

\bibitem{AloBere10}
R. Aloisio, V. Berezinsky and S. Grigorieva, arXiv:1006.2484 [astro-ph.CO]; 
R. Aloisio, V. Berezinsky and S. Grigorieva, arXiv:0802.4452 [astro-ph].

\bibitem{Stecker_EBL}
F.W. Stecker, M.A. Malkan and S.T. Scully, ApJ {\bf 648} (2006) 774.

\bibitem{BereE1/2}
V. Berezinsky and S. Grigorieva, Astron. Astrophys. {\bf 199} (1988) 1-12.

\bibitem{disapp}
R. Aloisio, V. Berezinsky and A. Gazizov, Astropart. Phys. {\bf 34} (2011) 620-626.

\end{thebibliography}

\end{document}